\documentclass[a4paper, 11pt]{article}
\usepackage{graphicx}
\usepackage{amsmath}
\usepackage{amsfonts}
\usepackage{natbib}
\usepackage{amssymb}
\usepackage{rotating}

\usepackage{setspace}

\newcommand{\ex}[1]{\ensuremath{\mathbb{E}[#1]}}

\newcommand{\corr}[1]{\ensuremath{\mathrm{Corr}[#1]}}
\newcommand{\bd}[1]{\ensuremath{\mbox{\boldmath $#1$}}}

\begin{document}

\doublespacing

\title{Controlling for unmeasured confounding and spatial misalignment in long-term air pollution and health studies}

\author{Duncan Lee$^1$ and Christophe Sarran$^2$\\
$^1$ School of Mathematics and Statistics, University of Glasgow\\
$^2$ UK Met Office, Exeter, UK}

\maketitle

\begin{abstract}
{The health impact of long-term exposure to air pollution is now routinely estimated using spatial ecological studies, due to the recent widespread availability of spatial  referenced pollution and disease data. However, this areal unit study design  presents a number of statistical challenges, which if ignored have the potential to bias the estimated pollution-health relationship. One such challenge is how to control for the spatial autocorrelation present in the data after accounting for the known covariates, which is caused by unmeasured confounding. A second challenge is how to adjust the functional form of the model to account for the spatial misalignment between the pollution and disease data, which causes within-area variation in the pollution data. These challenges have largely been ignored in existing long-term spatial air pollution and health studies, so here we propose a novel Bayesian hierarchical model that addresses both challenges, and provide software to allow others to apply our model to their own data. The effectiveness of the proposed model is compared by simulation against a number of state of the art alternatives proposed in the literature, and is then used to estimate the  impact of nitrogen dioxide and particulate matter concentrations on respiratory hospital admissions in a new epidemiological study in England in 2010 at the Local Authority level. }

\textbf{Air pollution and health; Spatial confounding; Spatial misalignment.}
\end{abstract}

\section{Introduction}
The health effects of air pollution came to public prominence in the mid 1900's, as a result of high-pollution episodes in the Meuse Valley in Belgium, Donora, Pennsylvania, and London, England. The latter occurred from 4th - 10th December 1952, and is estimated to have resulted in around 12,000 excess deaths compared with preceding years (\citealp{bell2001}). Although air pollution concentrations are now much lower than in the mid 1900's in much of the world, air pollution remains a global public health problem. For example, in April 2014 the World Health Organisation  estimated that outdoor air pollution was responsible for the deaths of 3.7 million people under the age of 60 in 2012. These health effect estimates are based on evidence from a large number of epidemiological studies, which collectively have estimated the health impact of exposure in both the short and the long term.  The health impact of long-term exposure has typically been estimated using cohort studies (see \citealp{cesaroni2014} and \citealp{dockery1993}), but the long-term follow up period required for the cohort makes them time consuming and expensive to implement. Therefore,  spatial ecological study designs have recently been used to quantify the health impact of long-term exposure to air pollution, and examples include  \cite{elliott2007} and \cite{lee2009}. These spatial ecological studies are inexpensive and quick to implement, due to the now routine availability of the required data. Thus, while they cannot provide individual-level evidence  on cause and effect, they independently corroborate  the body of evidence provided by cohort studies about the long-term impact of air pollution.\\

Spatial ecological studies are based on geographical contrasts in both air pollution concentrations and disease risks, over a set of non-overlapping areal units. The disease data take the form of aggregated counts of adverse health events, meaning that Poisson log-linear models are typically used for the analysis. The spatial pattern in disease risk is modelled by known covariates and a set of spatially autocorrelated random effects,  the former including air pollution concentrations and measures of  socio-economic deprivation and demography. The network of air pollution monitors is typically sparse relative to the number of areal units at which disease counts are measured, so modelled concentrations on a regular grid from atmospheric dispersion models are commonly used instead (see \citealp{lee2009} and \citealp{haining2010}). The  random effects account for any spatial autocorrelation and overdispersion (with respect to the Poisson likelihood) remaining in the disease data after the covariate effects have been accounted for, which could be caused by unmeasured confounding,  neighbourhood effects and grouping effects. This spatial autocorrelation is typically allowed for by modelling the random effects with a conditional autoregressive (CAR, \cite{besag1991}) prior distribution, as part of a hierarchical Bayesian model. A relatively small number of spatial ecological studies have been published to date (for example \citealp{jerrett2005}, \citealp{maheswaran2005}, \citealp{elliott2007}, \citealp{janes2007}, \citealp{lee2009},  \citealp{young2009}, \citealp{greven2011}, \citealp{lawson2012}), but the majority suffer from potential statistical limitations that could bias the estimated health risks.\\

The first potential limitation of many existing models is their approach to dealing with the residual spatial autocorrelation in the data after the known covariate effects have been accounted for.  Existing research  by \cite{clayton1993} has shown the potential for collinearity between spatially smooth random effects and any covariate in the model that is also spatially smooth, such as air pollution. This potential collinearity can lead to variance inflation and instability in the estimation of the air pollution effect. Additionally, the residual autocorrelation in the disease data is not necessarily globally spatially smooth, because the disease data  are unlikely to be globally smooth and part of any smooth variation has been removed by including air pollution in the model. Instead, the residual autocorrelation is likely to be localised, and be strong between
some pairs of adjacent areal units whilst other adjacent pairs exhibit very different residual values. Thus global smoothing models are insufficiently flexible to capture this localised spatial autocorrelation. The second statistical limitation arises because the pollution and disease data are spatially misaligned, as the modelled concentrations are available on a regular grid while the disease counts relate to irregularly shaped administrative units. The modelled pollution concentrations are typically at a finer spatial scale than the disease counts, resulting in variation in the pollution concentrations within an areal unit. This within-area variation in concentrations has been ignored by the majority of existing studies which compute the average concentration in each areal unit, and only \cite{haining2010} have allowed for such variation in this setting. \\

Therefore this paper makes two important contributions to the literature on spatial air pollution epidemiology. Firstly, it presents a novel overarching Bayesian model for spatial air pollution and health studies, which is the first of its type to simultaneously address the dual problems of capturing localised residual spatial autocorrelation and spatial misalignment between the pollution and disease data. We also provide software to implement our overarching model, via functionality in the \emph{CARBayes} software package for the statistical software R (\citealp{R}).  Secondly, we present the first comprehensive comparative assessment of the range of state of the art  models used in the literature, and quantify the likely nature and size of any bias in their estimated pollution-health relationships. The statistical modelling issues are motivated by a new study quantifying the effect of air pollution concentrations on respiratory ill health in England in 2010, and the data for this study are presented in Section 2. Section 3 outlines the statistical modelling challenges in such studies, summarises the methods that have been proposed, and describes the overarching model for which software is available. Section 4 quantifies the impact of inappropriate statistical modelling on the estimated air pollution effect, while the results of the England study are presented in Section 5. Finally, Section 6 concludes the paper with areas for future work.

\section{Data description}
\begin{figure}
\centering\caption{The top left panel displays the standardised morbidity ratio for hospital admissions due to respiratory disease in 2010, while the top right panel presents the modelled annual mean PM$_{2.5}$ concentrations at a 1 kilometre square resolution. The bottom left panel displays the number of modelled pollution concentrations in each LUA, while the bottom right panel displays the residuals from fitting a simple Poisson generalised linear model to the data.}\label{data1}
\begin{picture}(10, 20)
\put(-4.5,7.5){\scalebox{0.75}{\includegraphics{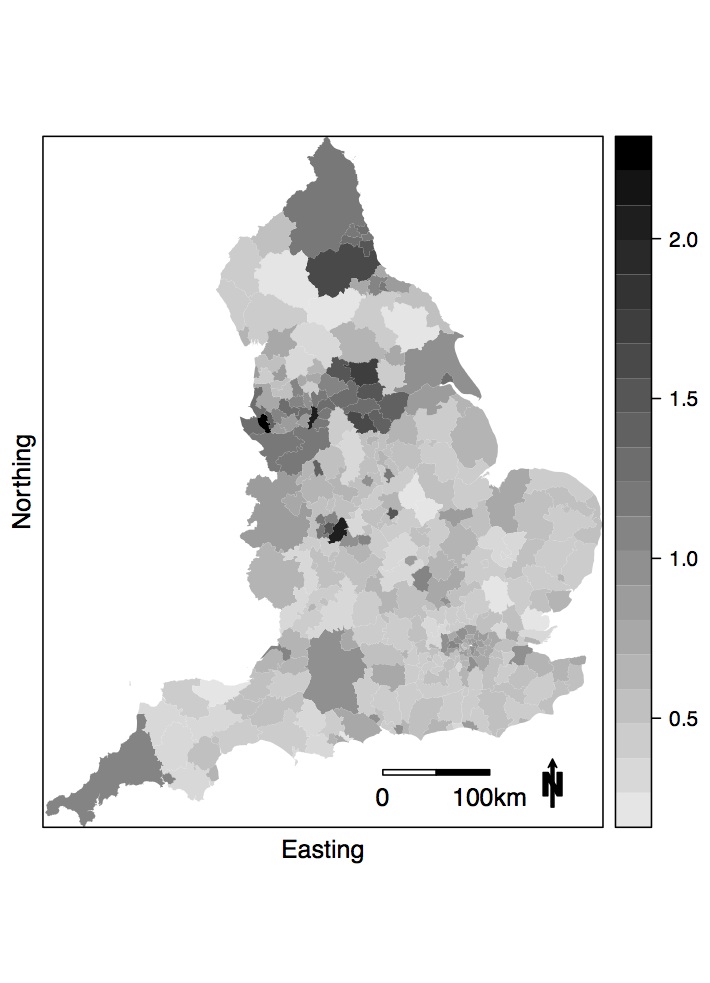}}}
\put(5,7.5){\scalebox{0.75}{\includegraphics{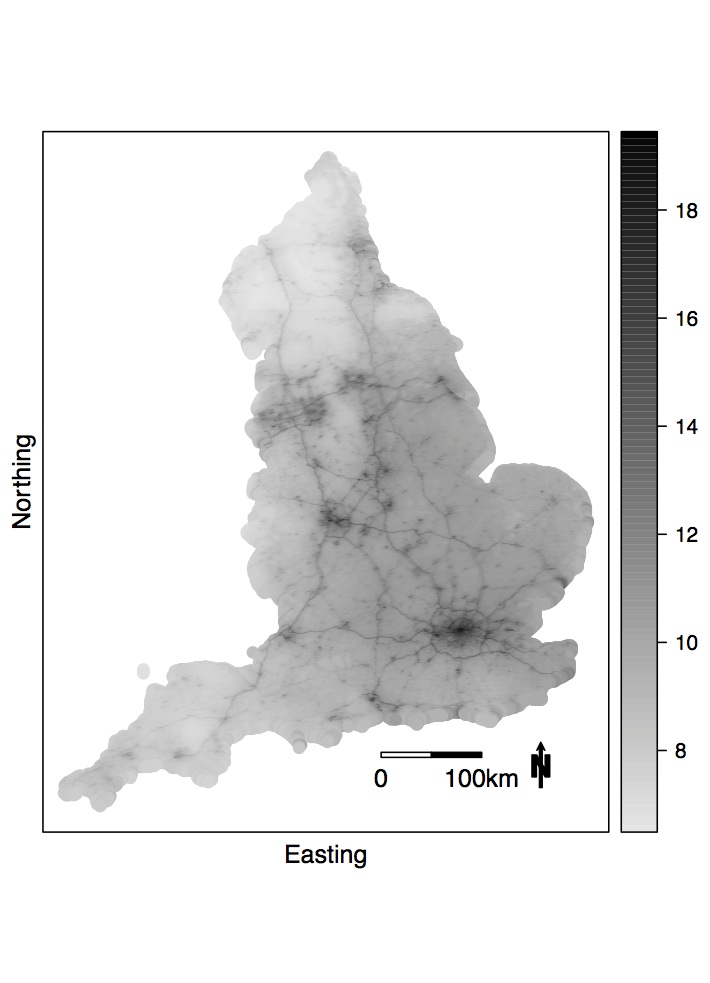}}}
\put(-4.5,-3.5){\scalebox{0.75}{\includegraphics{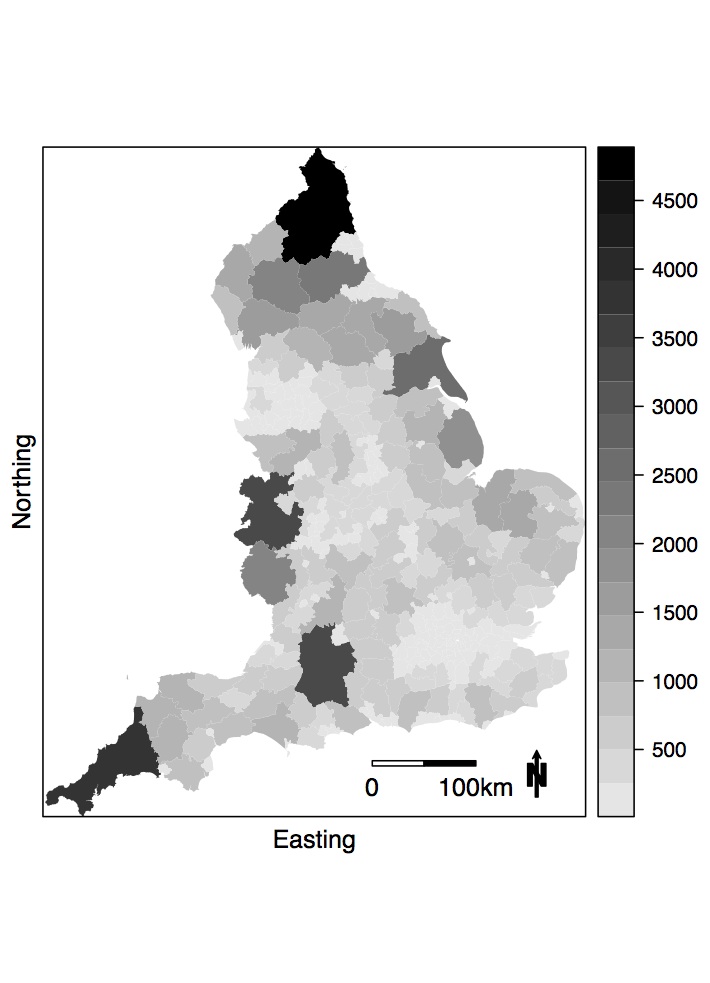}}}
\put(5,-3.5){\scalebox{0.75}{\includegraphics{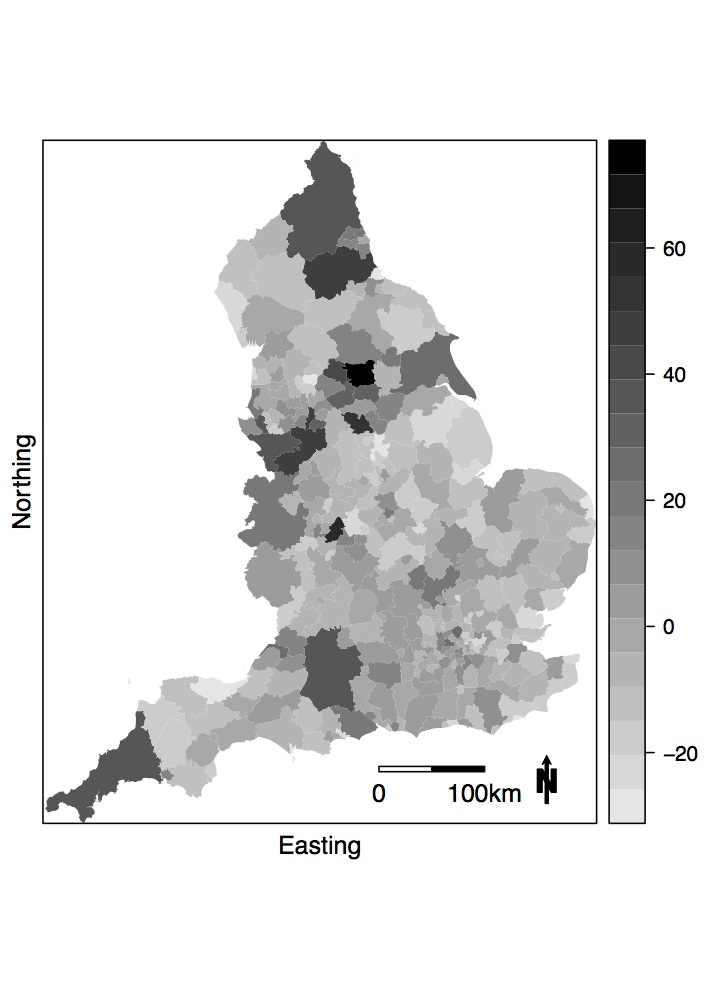}}}
\end{picture}
\end{figure}

The model developed in this paper is motivated by a new study investigating the health impact of long-term exposure to air pollution in England in 2010, which had a population of 52.6 million people. Hospital admission records from the Health and Social Care Information Centre (\emph{www.hscic.gov.uk}) were analysed at the UK Met Office to provide counts of hospital admissions by local authority, where the primary diagnosis was circulatory or respiratory disease and where the method of admission was as an emergency. The resulting disease data are aggregated counts of the numbers of hospital admissions from respiratory disease (International Classification of Disease tenth revision  codes J00-J99) in 2010 across the $n=323$ Local and Unitary Authorities (LUA) that make up mainland England. Differences in the population sizes and demographic structures between areal units are accounted for by computing the expected number of hospital admissions for each  LUA based on national age and sex specific admission rates. Based on these data an exploratory measure of disease risk is the Standardised Morbidity Ratio (SMR), which for area $k$ is the ratio of the observed to the expected numbers of disease cases. The SMR is summarised in the top left panel of Figure \ref{data1} and Table \ref{data2}, the former displaying the SMR as a choropleth map. The figure shows that the highest risk areas are cities in the north and central parts of England, such as Liverpool, Birmingham and Manchester. In contrast, the lowest risk areas are typically rural, and include West Somerset in the far south west of the country. The SMR map shows evidence of localised spatial smoothness, with some pairs of neighbouring areal units exhibiting similar risks while other pairs have vastly different values.\\

Modelled yearly average concentrations of a number of pollutants at a resolution of 1 kilometre grid squares are available for this study, and were produced by \cite{AEA2011} and are available to download from \emph{http://uk-air.defra.gov.uk/data/pcm-data}. The pollutants we consider in this study are nitrogen dioxide (NO$_{2}$) and particulate matter, the latter being measured as particles less than 2.5$\mu m^{-3}$ (PM$_{2.5}$) and 10$\mu m^{-3}$ (PM$_{10}$) respectively. These modelled background concentrations have complete spatial coverage of England, and the top right panel of Figure \ref{data1} displays the PM$_{2.5}$ concentrations from which the major cities and motorway network can be clearly seen.  These data are thus spatially misaligned to the respiratory disease data, and the bottom left panel of Figure  \ref{data1} displays the number of modelled pollution concentrations lying within each LUA. The figure shows that on average there are 215 pollution values for each LUA, with a  range from 11 to 4889. The distribution of means and coefficients of variation (standard deviation divided by the mean) in the 1 kilometre values within each of the 323 LUA are displayed in Table \ref{data2}, which shows that the greatest relative levels of within LUA variation in concentrations come from NO$_{2}$. The relationships between the mean and variance of each pollutant within a LUA are positive and linear, with correlation coefficients of  0.61 (NO$_{2}$), 0.48 (PM$_{2.5}$) and 0.16 (PM$_{10}$) respectively. These positive mean-variance relationships will determine the appropriate functional form of the pollution-health model proposed in Section 3.\\

Finally, socio-economic deprivation is known to have a major impact on disease risk, and here we control for it using two proxy measures. The first is the percentage of the working age population in each LUA that are in receipt of Job Seekers Allowance (JSA), which Table \ref{data2} shows ranges between 5.77$\%$ and 27.57$\%$. The second is the natural log of the average property price in each LUA, which on the original scale ranges between $\pounds 67,507$ and $\pounds 751,630$. A natural log transformation is used here because exploratory analyses showed it exhibited a stronger relationship with disease risk. These covariates were included in a Poisson generalised linear model along with PM$_{2.5}$ concentrations, and the residuals are displayed in the bottom right panel of Figure \ref{data1}. The residuals exhibit strong spatial autocorrelation, with a significant Moran's I statistic of 0.282 (p-value 0.00001). However, the autocorrelation is visually localised, with some pairs of neighbouring regions having very different residual values. Thus a model that accounts for this localised residual spatial autocorrelation is required, and such a model is proposed in Section 3.

\begin{table}
\small\caption{The table summarises the distribution of the hospital admissions (as a standardised morbidity ratio), covariate and pollution data over the $n=323$ LUA in England. The pollution summaries relate to the distribution of means and coefficients of variation (CoV) of the 1 kilometre  concentrations within each LUA.}\label{data2}
\centering\begin{tabular}{lrrrrr}
\hline
\raisebox{-1.5ex}[0pt]{\textbf{Variable}}&\multicolumn{5}{c}{\textbf{Distribution}}\\
&\textbf{0$\%$} &\textbf{25$\%$} &\textbf{50$\%$} & \textbf{75$\%$} &\textbf{100$\%$}  \\  \hline

SMR&0.17 &0.45 &0.57 &0.83 &2.31\\
Log house price ($\pounds$) &11.12 &11.90 &12.14 &12.34 &13.53\\
Jobs Seekers Allowance ($\%$)&5.77  &9.99 &13.29 &17.15 &27.57\\\hline

\textbf{Mean}&&&&&\\
NO$_{2}$&4.71 &11.84 &15.97 &20.75 &47.04\\
PM$_{2.5}$&6.87  &9.76 &10.76 &11.64 &16.74\\
PM$_{10}$&9.67 &14.41 &15.97 &17.20 &23.28\\\hline

\textbf{CoV}&&&&&\\
NO$_{2}$&  0.06& 0.16& 0.22& 0.27& 0.48\\
PM$_{2.5}$& 0.02& 0.05& 0.06& 0.08& 0.15\\
PM$_{10}$& 0.03 &0.06 &0.07 &0.08 &0.19\\\hline
\end{tabular}
\normalsize
\end{table}

\section{Modelling}
This section outlines a new Bayesian hierarchical model for estimating the health effects of air pollution in spatial ecological studies, which is the first of its type to simultaneously address the dual problems of, potentially localised, residual spatial autocorrelation in the disease data after accounting for known covariates,  and spatial misalignment between the pollution and disease data resulting in within-area variation in the pollution concentrations. The first part of this section presents the general likelihood model, while the second and third parts describe the spatial autocorrelation and pollution-health components of the model respectively. The model class proposed here is implemented in the \emph{R} software environment, via the freely available \emph{CARBayes} package.

\subsection{Level 1 - Likelihood model}
The study region is partitioned into $k=1,\ldots,n$ areal units, and the vectors of observed and expected numbers of disease cases  across the $n$ units are denoted by  $\mathbf{Y}=(Y_{1},\ldots,Y_{n})$ and $\mathbf{E}=(E_{1},\ldots,E_{n})$ respectively. The latter account for differences in the size and demographic structure of the populations living in each areal unit, and are computed by external standardisation using age and sex specific disease rates for the whole study region. Let $\mathbf{w}=\{\mathbf{w}_{1},\ldots,\mathbf{w}_{n}\}$ denote the set of modelled air pollution data for the entire study region, where $\mathbf{w}_{k}=(w_{k1},\ldots,w_{k q_{k}})$ is the vector of  pollution concentrations observed at $q_{k}$ different locations within the $k$th area unit. These data are used to estimate $\mathbf{W}=(W_{1},\ldots,W_{n})$, the vector of random variables representing the distributions of air pollution concentrations within each of the $n$ areal units at which disease data are observed. Finally, $\mathbf{X}=(\mathbf{x}_{1}^{\tiny\mbox{T}\normalsize},\ldots,\mathbf{x}_{n}^{\tiny\mbox{T}\normalsize})^{\tiny\mbox{T}\normalsize}$ is a matrix of $p$ known covariates including measures of deprivation and a column of ones for the intercept term, where $\mathbf{x}_{k}^{\tiny\mbox{T}\normalsize}=(1,x_{k2},\ldots,x_{kp})$ are the values for area $k$. Poisson log-linear models are typically used to estimate the health impact of air pollution as the disease data are counts, and here we propose the general specification:

\begin{eqnarray}
Y_{k}|E_k,R_{k} &\sim&\mbox{Poisson}(E_kR_k)~~~~\mbox{for }k=1,\ldots,n,\label{equation likelihood}\\
R_{k}&=&\exp(\mathbf{x}_{k}^{\tiny\mbox{T}\normalsize}\bd{\beta} + \phi_{k})g(W_{k}; \mathbf{w}_{k}, \alpha).\nonumber
\end{eqnarray}

The relative risk of disease in area $k$ is denoted by $R_{k}$, and $R_{k}=1.2$ corresponds to a 20$\%$ increased risk of disease compared with the expected disease cases $E_{k}$. The vector of covariate effects are denoted by $\bd{\beta}=(\beta_{1},\ldots,\beta_{p})$, and are typically assigned a multivariate Gaussian prior with weakly informative hyper parameters $(\bd{\mu}_{\bd{\beta}}, \Sigma_{\bd{\beta}})$. The two key modelling challenges for these data relate to the remaining model components, and are the focus of the remainder of this section. The first of these is the vector $\bd{\phi}=(\phi_1,\ldots,\phi_n)$, which accounts for any spatial autocorrelation  or overdispersion remaining in the data after the covariate effects have been accounted for, and is discussed in Section 3.2. The second is the focus of Section 3.3, and is the modelling mechanism $g(W_{k}; \mathbf{w}_{k}, \alpha)$ that relates the observed air pollution concentrations $\mathbf{w}_{k}$, which are realisations of the random variable $W_{k}$, to the disease counts, whilst accounting for the spatial misalignment of these data.

\subsection{Level 2 - Spatial autocorrelation model}
The vector of random effects $\bd{\phi}$ is typically assumed to be a globally spatially smooth process, but \cite{clayton1993} and \cite{reich2006} have shown the potential for collinearity between $\bd{\phi}$ and any covariate in the model such as air pollution that is spatially smooth, leading to variance inflation and bias in the estimated covariate effects. This problem is likely to arise here, because yearly mean background pollution concentrations are spatially smooth after being averaged over  LUAs. Additionally, the residuals from fitting a simple Poisson generalised linear model to the England data are spatially autocorrelated (see Section 2), but Figure \ref{data1} shows this autocorrelation is localised and not a globally smooth surface.\\

Therefore we propose modelling $\bd{\phi}$ as locally and not globally smooth, so that two random effects in adjacent LUA can be either spatially autocorrelated or exhibit very different values, with the choice being determined by the data. We achieve this by partitioning $\bd{\phi}$ into a globally smooth surface and a piecewise constant intercept surface, which allows neighbouring areas $(k,i)$  to have very similar values ($\phi_{k}, \phi_{i}$) if they have the same intercept or very different if they have different intercepts. The piecewise constant intercept surface has at most $G$ distinct intercepts, $\bd{\lambda}=(\lambda_{1},\ldots,\lambda_{G})$, and the model we propose is given by:

\begin{eqnarray}
\phi_{k}& = &\lambda_{Z_{k}} + \theta_{k},\label{equation localCAR}\\
\lambda_i & \sim & \mbox{Uniform}(\lambda_{i-1}, \lambda_{i+1})\hspace{1cm}\mbox{for }i=1,\ldots,G,\nonumber\\
f(Z_{k})&=&\frac{\exp(-\delta(Z_{k}-G^{*})^{2})}{\sum_{r=1}^{G}\exp(-\delta(r-G^{*})^{2})},\nonumber\\
\delta&\sim&\mbox{Uniform}(0, M=100),\nonumber\\
\theta_{k}| \bd{\theta}_{-k}, \tau^{2}, \rho,\mathbf{W}&\sim&\mbox{N}\left(\frac{\rho\sum_{i=1}^{n}w_{ki}\theta_{i}}{\rho\sum_{k=1}^{n}w_{ki} + 1-\rho},~
\frac{\tau^{2}}{\rho\sum_{k=1}^{n}w_{ki} + 1-\rho}\right),\nonumber\\
\tau^{2}&\sim&\mbox{Inverse-gamma}(a=0.001,b=0.001),\nonumber\\
\rho&\sim&\mbox{Uniform}(0, 1).\nonumber
\end{eqnarray}

The piecewise constant intercept term $\bd{\lambda}$ is ordered as $\lambda_{1}<\lambda_{2}<\ldots<\lambda_{G}$ to mitigate against the label switching problem common in mixture models, and in (\ref{equation localCAR}) $\lambda_{0}=-\infty$ and $\lambda_{G+1}=\infty$. Area $k$ is assigned to one of the $G$ intercepts by $Z_{k}\in\{1,\ldots,G\}$, and here we fix the number of different intercepts $G$ to be overly large and penalise $Z_{k}$ towards the middle intercept value. This is achieved by the penalty term $\delta(Z_{k}-G^{*})^{2}$ in the prior for $Z_{k}$, where $G^{*}=(G+1)/2$ is the middle of the intercept terms. A weakly informative uniform prior with a large range is specified for the penalty parameter $\delta$, so that the data play the dominant role in estimating its value. The second component in $\bd{\phi}$, $\bd{\theta}=(\theta_{1},\ldots,\theta_{n})$ captures spatially smooth structure, and is modelled by the Conditional autoregressive (CAR, \citealp{besag1991}) prior proposed by \cite{leroux1999}. Spatial autocorrelation is induced into these random effects via a binary $n\times n$ neighbourhood matrix $\mathbf{W}$, where $w_{ki}=1$ if areal units $(k,i)$ share a common border (denoted $k\sim i$) and $w_{ki}=0$  otherwise (denoted $k\nsim i$). CAR priors are commonly specified as full conditional distributions, $f(\theta_{k}|\bd{\theta}_{-k})$ for $k=1,\ldots,n$, where $\bd{\theta}_{-k}=(\theta_{1},\ldots,\theta_{k-1},\theta_{k+1},\ldots,\theta_{n})$, and are a special case of a Gaussian Markov Random Field (GMRF). In (\ref{equation localCAR}) $\rho$ is a global spatial smoothness parameter, where $\rho=1$ corresponds to the intrinsic CAR prior for strong spatial smoothing proposed by \cite{besag1991}, while $\rho=0$, corresponds to independent  random effects.  The global level of spatial autocorrelation imposed by this model can be seen from its implied partial autocorrelation between $(\theta_{k}, \theta_{i})$, which is given by:

\begin{equation}
\corr{\theta_{k},\theta_{i}|\bd{\theta}_{-ki}}~=~\frac{\rho w_{ki}}{\sqrt{(\rho\sum_{j=1}^{n}w_{kj} + 1-\rho)(\rho\sum_{l=1}^{n}w_{il} + 1-\rho)}}\label{equation partialcorrelation}.
\end{equation}

Here, for all pairs of adjacent areal units $(k,i)$, $w_{ki}=1$, and hence they are partially autocorrelated with the level of that autocorrelation controlled globally for all adjacent pairs by $\rho$.\\

A number of alternatives have been proposed for modelling residual spatial autocorrelation in existing spatial ecological studies, and we provide a brief review here and compare their relative performances in a simulation study in Section 4. The simplest approach is to ignore the spatial autocorrelation and set $\phi_k=0$, but the most common approach is to use a CAR model such as that given by (\ref{equation localCAR}) with the piecewise constant  intercept term removed (so that $\phi_{k}=\theta_{k}$). This approach has been adopted by  \cite{maheswaran2005} and \cite{lee2009}, and has been shown by \cite{reich2006} to have the potential for the collinearity problems discussed above. Therefore recently \cite{hughes2013}  proposed representing  $\bd{\phi}$ with a set of basis functions that are orthogonal to the covariates and are spatially smooth, using the Eigenvectors from the matrix product $\mathbf{PWP}$, where $\mathbf{P}$ is the residual projection matrix from fitting a normal linear model to the data. However, this model cannot capture localised spatial structure,  and an alternative approach is localised CAR smoothing, with examples including \cite{lu2007}, \cite{lee2013} and \cite{lee2014}. This approach treats the elements of the neighbourhood matrix $\mathbf{W}$ corresponding to adjacent areal units as random quantities to be estimated in the model, rather than fixed at one. Thus if $w_{ki}=1$ the  corresponding random effects are partially autocorrelated inducing smoothness between them (see (\ref{equation partialcorrelation})), while if $w_{ki}=0$ they are conditionally independent and no such smoothing is induced. \cite{lee2013} propose an iterative algorithm for estimation, in which $\mathbf{W}$ is updated deterministically based on the joint posterior distribution of the remaining model parameters.

\subsection{Level 3 - The pollution-health relationship $g(W_{k}; \mathbf{w}_{k}, \alpha)$}
The modelled  pollution concentrations are spatially misaligned with the disease data, as the former are available at a 1 kilometre square resolution while the latter are population summaries for LUA that are irregular in shape and larger than a 1 kilometre square. Therefore, areal unit $k$ has  $q_{k}$ concentrations $\mathbf{w}_{k}=(w_{k1},\ldots,w_{k q_{k}})$, and hence there is variation in the pollution concentrations within an areal unit for which only a single disease count exists. Furthermore, the population may be unevenly distributed within each areal unit, and the proportions of the population in each of the $q_{k}$ grid squares within areal unit $k$ are denoted by $\mathbf{p}_{k}=(p_{k1},\ldots,p_{k q_{k}})$, where $\sum_{i=1}^{q_{k}}p_{ki}=1$. The standard model for relating air pollution to health in the existing literature has the functional form

\begin{equation}
R_{k}~=~\exp(\mathbf{x}_{k}^{\tiny\mbox{T}\normalsize}\tilde{\bd{\beta}} + \tilde{\phi}_{k} + \mu_{k}\tilde{\alpha}),\label{equation ecological}
\end{equation}

where $\mu_{k}=\ex{W_{k}}$ is the average pollution concentration in areal unit $k$ and can be estimated from the observed data as

\begin{equation}
\hat{\mu}_{k}~=~\sum_{i=1}^{q_{k}}p_{ki}w_{ki},
\label{equation popweighted}
\end{equation}

the population weighted average concentration within areal unit $k$. However, \cite{wakefield2001} and \cite{wakefield2006} have shown it is inappropriate to  average the pollution concentrations via (\ref{equation ecological}) and (\ref{equation popweighted}), as this ignores the spatial variation in the concentrations within an areal unit. Specifically, let $Y_{ki}$ for $i=1,\ldots,q_{k}$,  denote the unknown number of disease cases from the  $p_{ki}$ proportion of the population  in areal unit $k$ that experienced pollution concentration $w_{ki}$. Then an appropriate model for the unobserved $Y_{ki}$ would be $Y_{ki}\sim\mbox{Poisson}(E_{k}p_{ki}R_{ki})$, where $R_{ki}=\exp(\mathbf{x}_{k}^{\tiny\mbox{T}\normalsize}\bd{\beta} + \phi_{k} + w_{ki}\alpha)$ and $\mathbf{x}_{k}^{\tiny\mbox{T}\normalsize}\bd{\alpha} + \phi_{k}$ are measured and unmeasured covariates that have the same effect across all $q_{k}$ sub-populations within areal unit $k$. Then as the observed disease data are $Y_{k}=\sum_{i=1}^{q_{k}}Y_{ik}$, we have that  $Y_{k}\sim\mbox{Poisson}\left(E_{k}R_{k}\right)$ where $R_{k}=\sum_{i=1}^{q_{k}}p_{ki}R_{ki}$, which simplifies to:

\begin{equation}
R_{k}~=~\exp(\mathbf{x}_{k}^{\tiny\mbox{T}\normalsize}\bd{\beta} + \phi_{k})\sum_{i=1}^{q_{k}}p_{ki}\exp(w_{ki}\alpha)\label{equation aggregate}.
\end{equation}

This means that $g(W_{k}; \mathbf{w}_{k},\alpha)=\sum_{i=1}^{q_{k}}p_{ki}\exp(w_{ki}\alpha)$ in the general model (\ref{equation likelihood}), which is similar to equation (3.3) in \cite{wakefield2006}. In the above equation $\sum_{i=1}^{q_{k}}p_{ki}\exp(w_{ki}\alpha)$ is the sample population weighted moment generating function for the random variable $W_{k}$ representing the distribution of pollution concentrations within the $k$th local authority, that is $\mbox{MGT}_{W_k}(\alpha)=\ex{\exp(W_k\alpha)}$. Thus, comparing (\ref{equation ecological}) to  (\ref{equation aggregate}) shows that within area variation in concentrations may cause bias in estimating $\alpha$ as $\ex{\exp(W_k\alpha)}\neq \exp(\ex{W_{k}}\alpha)$, where $\ex{W_{k}}=\mu_{k}$. \\

In this paper we use the sample representation (\ref{equation aggregate}) of the moment generating function when estimating the health effects of pollution, but the magnitude of the potential bias can in part be predicted depending on both the distribution of the within-area variation of concentrations assumed for $W_{k}$ and the relationship between its mean and variance. For example, assume that population density is constant ($p_{ki}=1/q_{k}$) and that $W_{k}\sim\mbox{N}(\mu_{k}, \sigma^{2}_{k})$, then the term $\sum_{i=1}^{q_{k}}p_{ki}\exp(w_{ki}\alpha)$ in (\ref{equation aggregate}) can be replaced by the moment generating function of $W_{k}$,  which is $\exp(\mu_{k}\alpha + 0.5\alpha^2\sigma^2_k)$. If the variance of $W_{k}$ increases with the mean in a linear fashion, as is approximately the case with the data presented in Section 2, then $\sigma^{2}_{k}=a+b\mu_{k}$ where $b>0$. Then comparing the multipliers of $\mu_{k}$ in (\ref{equation ecological}) and $\exp(\mu_{k}\alpha + 0.5\alpha^2\sigma^2_k)$  (with $\sigma^{2}_{k}=a+b\mu_{k}$) shows that $\tilde{\alpha}=\alpha + 0.5b\alpha^{2}$. Thus there will be no bias under this setting if the mean and variance are independent ($b=0$), while for small $\alpha$ the bias should be small due to the $\alpha^{2}$ in the bias term $0.5b\alpha^{2}$.

\section{Simulation study}
This section presents two simulation studies, which respectively assess the impact that different types of residual spatial confounding and within-area variation in the pollution concentrations have on health effects  estimation. In the first study the localised smoothing model (\ref{equation localCAR}) proposed here is compared against the alternative approaches to dealing with spatial confounding discussed in the last paragraph of Section 3.2, while in the second study the aggregate model (\ref{equation aggregate}) is compared against the naive ecological model (\ref{equation ecological}). 

\subsection{Study 1 - Spatial confounding}

\subsubsection{Data generation and study design}
Simulated data are generated for the $n=323$ LUA comprising mainland England, which is the study region for the motivating study described in Section 2. Disease counts are generated from a Poisson$(E_{k}R_{k})$ model, where the expected counts $E_{k}$ are generated from a uniform distribution on the range $[70, 130]$ giving a moderate disease prevalence in terms of the existing literature. The log-risk surface is generated as a linear combination of a spatially smooth covariate acting as air pollution and residual spatial autocorrelation. For this first study the pollution covariate is assumed to have no within area variation, and is generated from a Gaussian spatial processes with a mean of 20 and a spatially smooth variance matrix defined by the Mat\'{e}rn family of  autocorrelation functions. For the latter the  smoothness parameter equals 2.5 and the range parameter equals 75. A linear relationship between air pollution and health is assumed in generating the disease count data, and the regression parameter corresponds to a 5$\%$ increase in disease risk for a 2$\mu gm^{-3}$ increase in pollution concentrations, which is similar to that observed for the real data in Section 5.\\

Eight different scenarios are considered for the spatial confounding component $\bd{\phi}$,  which cover the range of scenarios likely to be seen in real data. The standard deviation of $\bd{\phi}$, denoted SD$_{\bd{\phi}}$, quantifying its level of variation is fixed at either 0.1 or 0.01, with the former obviously likely to have a larger effect on the fixed effects estimates than the latter. Four different spatial structures are considered for $\bd{\phi}$, which are: (A) independent in space; (B) globally spatially smooth with a smaller range parameter (less smooth) than the pollution covariate; (C) globally spatially smooth with  the same range as the pollution covariate; and (D) locally smooth with high autocorrelation between some pairs of adjacent areas and step changes between other pairs. Example realisations of these spatial surfaces are displayed in Figure 1 of the supplementary material accompanying this paper.   We apply five different models to each simulated data set, which include the localised smoothing model (\ref{equation localCAR}) proposed here (denoted \textbf{Model-Local}) with $G=5$, as well as  a simple over dispersed Poisson log-linear model (denoted \textbf{Model-GLM}), a spatially smooth CAR model (denoted \textbf{Model-CAR}) and the models of \cite{hughes2013} (denoted \textbf{Model-HH}) for orthogonal autocorrelation and \cite{lee2013} (denoted \textbf{Model-LM}) for localised autocorrelation.

\subsubsection{Results}
Five hundred data sets are generated under each of the 8 scenarios described above, and the percentage bias and percentage root mean square error (RMSE) of the estimated regression parameter for air pollution  are computed as $\mbox{Bias}(\alpha)=100\times(\frac{1}{500}\sum_{i=1}^{500}(\hat{\alpha}_{i} - \alpha))/\alpha$ and $\mbox{RMSE}(\alpha)=100\times(\sqrt{\frac{1}{500}\sum_{i=1}^{500}(\hat{\alpha}_{i} - \alpha)^{2}})/\alpha$ respectively, where $\hat{\alpha}_{i}$ is the estimate for the $i$th simulated data set from each model. Also computed are the coverage probabilities of the 95$\%$ uncertainty intervals for the estimated air pollution effect, and all the results are displayed in Table \ref{table simulation1}. None of the models display any systematic bias for any of the scenarios, with biases below $1\%$ in nearly all cases. 
Scenario A corresponds to no spatial confounding (the residual surface $\bd{\phi}$ is independent in space), and all models perform similarly with less than $7\%$ RMSE and coverages close to 95$\%$. The only exception to this is for \textbf{Model-HH} when the spatial confounding standard deviation is 0.1, whose coverage is less than 85$\%$. This relatively poor coverage for \textbf{Model-HH} is consistently observed for many of the remaining scenarios (especially when the spatial confounding standard deviation is 0.1), and suggests that it is not an appropriate model to use in this context.\\

Scenarios B and C correspond to increasing spatial confounding, and all models exhibit dramatic rises in RMSE and falls in coverage with the larger spatial confounding standard deviation of 0.1. \textbf{Model-CAR}, \textbf{Model--Local} and \textbf{Model-LM} perform the best in this regard and have similar results, in terms of both RMSE and coverage. Surprisingly \textbf{Model-HH}, which is designed to overcome this spatial confounding, does not outperform the other models that are prone to suffer from this confounding. When the standard deviation of $\bd{\phi}$ drops to 0.01 the results are similar to those under scenario A, because the level of spatial confounding is very small and hardly affects the fixed effects estimates. Finally, under the localised spatial confounding scenario D \textbf{Model--Local} performs best followed by \textbf{Model-LM} in terms of RMSE, and have lower values than the remaining models by up to 10 times. This is because they are the only models that allow for a non-constant level of spatial smoothing across the study region, and are specifically designed to allow for localised smoothness. However, \textbf{Model--Local} performs worse in terms of coverage than the globally smooth CAR model (\textbf{Model-CAR}) when SD$_{\bd{\phi}}$=0.1, which is due to an underestimation of the random effects variance caused by the piecewise intercept component $\lambda_{Z_{k}}$ modelling a proportion of the variation in the data. In contrast, the random effects in \textbf{Model-CAR} are modelling both the jumps and the smooth variation, and hence have a larger estimated variance.

\begin{table}
\caption{\label{table simulation1} The top panel displays the bias (as a $\%$ of the true value) for the pollution-health relationship estimated by each of the six models, the middle panel displays the root mean square error (as a $\%$ of the true value), while the bottom panel displays the coverage probabilities (as a $\%$) of the 95$\%$ uncertainty intervals.}
\small\centering\begin{tabular}{llrrrrr}
\hline
\raisebox{-1.5ex}[0pt]{\textbf{Scenario}}&\raisebox{-1.5ex}[0pt]{SD$_{\bd{\phi}}$}&\multicolumn{5}{c}{\textbf{Model}}\\
&&\textbf{Model-GLM}&\textbf{Model-CAR}&\textbf{Model-Local}&\textbf{Model-HH}&\textbf{Model-LM}\\\hline
&&&&&&\\
\textbf{Bias}&&&&&&\\\hline
A&0.1&0.04&    0.06&   -0.01&    0.08&    0.06 \\
&0.01&   0.24&    0.23&    0.23&    0.28&    0.23 \\
B&0.1&0.03&    0.06&    0.09&    0.07&    0.05 \\
&0.01&  -0.59&   -0.59&   -0.60&   -0.59&   -0.61\\
C&0.1&1.10&    0.60&    0.62&    0.83&    0.66 \\
&0.01&-0.44&   -0.43&   -0.43&   -0.40&   -0.41\\
D&0.1&  -0.29&   -0.93&   -0.25&   -0.22&    0.26 \\
&0.01&0.86&    0.01&    0.13&   -0.06&    1.94\\\hline

&&&&&&\\
\textbf{RMSE}&&&&&&\\\hline
A&0.1&6.55&6.52&6.67&6.56&6.61\\
&0.01&4.78&4.80&4.78&4.78&4.80\\
B&0.1&24.07&22.41&22.11&24.00&22.77\\
&0.01&4.88&4.88&4.88&4.88&4.89\\
C&0.1&35.16&28.10&28.26&35.09&27.44\\
&0.01&5.65&5.65&5.63&5.66&5.63\\
D&0.1&59.76&50.54&23.84&53.19&28.48\\
&0.01&58.76&44.91&4.78&49.21&10.73\\\hline

&&&&&&\\
\textbf{Coverage}&&&&&&\\\hline
A&0.1&94.6&97.0&94.8&84.6&97.4\\
&0.01&93.2&96.0&94.4&94.6&96.2\\
B&0.1&41.0&75.8&76.2&31.2&78.8\\
&0.01&95.0&96.8&96.8&95.6&97.0\\
C&0.1&22.2&62.0&60.8&17.8&63.8\\
&0.01&89.0&91.6&91.4&90.2&93.0\\
D&0.1&62.0&89.0&60.4&17.2&79.0\\
&0.01&58.2&95.4&94.4&15.2&95.6\\\hline

\end{tabular}
\normalsize
\end{table}

\subsection{Study 2 results - Within-area variation in pollution}

\subsubsection{Data generation and study design}
Simulated data for this study are generated using the same approach to that taken in the previous study, except that the pollution concentrations are assumed to vary within each LUA. The number of concentrations observed for each LUA is the same as in the real data, and we examine the impact that this variation has on the estimated health effects under a number of different scenarios. Specifically we vary the size of the estimated health effects and the type of within-area variation in pollution, as the discussion in Section 3 suggests that both will affect the results. The pollution-health relationships considered here have relative risks of 1.05 and 1.5 for a 2$\mu gm^{-3}$ increase in concentrations, as the former is similar to that seen in the England data analysed in Section 5 while the latter is chosen to be overly large. Both the absolute level of within-area variation and its relationship with the mean pollution level are varied in this study, with the former having standard deviations of 1 and 10 while the mean and variance of the pollution concentrations are assumed to be either unrelated or positively linearly related.  In all cases the within area variation in concentrations is generated from a Gaussian distribution. In all scenarios we compare the  aggregate model (\ref{equation aggregate}) proposed here with the naive ecological model (\ref{equation ecological}) commonly used in these studies, and both are completed by (\ref{equation likelihood}) and (\ref{equation localCAR}). 

\subsubsection{Results}
Five hundred data sets are generated under each of the 8 scenarios described above, and the results presented in Table \ref{table simulation2} include the percentage bias, percentage RMSE and coverage probabilities for the estimated pollution-health relationships. The top panel of the table shows the results for a realistic relative risk of 1.05 for a 2$\mu gm^{-3}$ increase in pollution, for different levels of within-area variation in the pollution concentrations. The table shows that neither the ecological model nor the aggregate model exhibit any systematic bias regardless of the level of within-area variation in the pollution concentrations, which is because the bias term $0.5b\alpha^{2}$ (based on the Gaussian and linearity assumptions, see Section 3.3) in the naive ecological model is small as the true value of $\alpha=0.0244$. The RMSE values and coverage probabilities for both models are similar, with the latter being close to their nominal 95$\%$ levels. These results thus suggest that while the ecological model is inappropriate mathematically, the small effect sizes seen in air pollution and health studies render its bias negligible in practice in this context. Conversely, for the larger effect size of a relative risk of 1.5 for a 2$\mu gm^{-3}$ increase in pollution ($\alpha=0.203$), the ecological model shows large bias and RMSE and low coverage if the within area variation in the pollution concentrations increases with the mean in a linear fashion. This result conforms to our theoretical expectations discussed in Section 3.3, while the aggregate model does not suffer from these problems. We note that the larger bias and RMSE observed when SD=1 than when SD=10 for the ecological model is at first surprising. However, as the SD increases so does the average size of the component $\sum_{i=1}^{q_{k}}p_{ki}\exp(w_{ki}\alpha)$, leading to larger counts $Y$ and hence the estimation accuracy improves. Finally, as expected, if the within-area variation in the pollution concentrations is independent of the mean pollution level, then the ecological model is once more unaffected.

\begin{table}
\small\caption{\label{table simulation2} The table displays the bias, RMSE (both as a $\%$ of the true value) and the coverage probabilities for the pollution-health relationship estimated by the naive ecological model (\ref{equation ecological}) and the aggregate model (\ref{equation aggregate}), as both the true risk and the within-area variation in pollution varies.}
\centering\begin{tabular}{llrrrrrr}
\hline
\raisebox{-1.5ex}[0pt]{\textbf{Risk ($\alpha$)}}&\raisebox{-1.5ex}[0pt]{\textbf{Pollution}}&\multicolumn{2}{c}{\textbf{Bias}}&\multicolumn{2}{c}{\textbf{RMSE}}&\multicolumn{2}{c}{\textbf{Coverage}}\\
&&\textbf{Model (\ref{equation ecological})}&\textbf{Model (\ref{equation aggregate})}&\textbf{Model (\ref{equation ecological})}&\textbf{Model (\ref{equation aggregate})}&\textbf{Model (\ref{equation ecological})}& \textbf{Model (\ref{equation aggregate})}\\\hline
1.05&SD=1, Indep&0.14&0.15&7.58&7.57&95.4&95.6\\
1.05&SD=1, Linear&0.23&0.17&8.45&8.44&93.8&94.0\\
1.05&SD=10, Indep&-0.07&-0.06&7.92&7.92&94.4&94.2\\
1.05 &SD=10, Linear&0.66&-0.01&8.26&8.12&93.6&93.8\\\hline
1.5&SD=1, Indep&024&-0.10&17.41&16.98&93.4&94.2\\
1.5&SD=1, Linear&19.58&-0.85&26.82&11.43&77.0&95.2\\
1.5&SD=10, Indep&-0.03&-0.26&7.56&4.95&95.0&96.8\\
1.5&SD=10, Linear&12.42&-0.10&16.27&2.25&75.8&96.2\\\hline
\end{tabular}
\normalsize
\end{table}

\section{Results from the England study}
\subsection{Modelling}
All models compared in the simulation studies were applied to the England data described in Section 2, which include the full aggregate model comprising (\ref{equation likelihood}), (\ref{equation localCAR}) and  (\ref{equation aggregate}) (\textbf{Model-Local-Agg})  proposed here (with $G=5$), and the naive ecological model comprising (\ref{equation likelihood}), (\ref{equation localCAR}) and (\ref{equation ecological})  (\textbf{Model-Local}, also with $G=5$) that ignores within-area variation in the pollution concentrations. These models allow a comparison of the effect of within-area variation in the pollution concentrations on health effects estimation for real data, and we also assess the impact of appropriately modelling residual spatial autocorrelation in the disease counts. This is achieved by comparing the above models to two commonly used in the literature, namely a simple generalised linear model (\textbf{Model-GLM}) and the extension including a set of spatially autocorrelated random effects  with a globally smooth conditional autoregressive prior (\textbf{Model-CAR}, model (\ref{equation localCAR}) with $\phi_{k}=\theta_{k}$). Additionally, we also compare our results to the recent models of \cite{hughes2013} (\textbf{Model-HH}) for orthogonal autocorrelation and \cite{lee2013} (\textbf{Model-LM}) for localised autocorrelation.\\

The covariates included in each model are one of the three pollutants considered here (NO$_2$, PM$_{2.5}$, PM$_{10}$) and the proxy measures of socio-economic deprivation, the latter including the percentage of the working age population in receipt of Job Seekers Allowance and the natural log of the median property price in each LUA. A single pollutant was included in each model due to the positive pairwise correlations between pollutants, which ranged between 0.79 and 0.94 at the  LUA  level. Such single pollutant analyses are the most common approach in the existing literature, with examples being  \cite{lee2009} and \cite{haining2010}. Total population counts were available for the 8546 Wards in England, which allows the pollution concentrations within each LUA to be population weighted. The models based on a single mean concentration in each LUA used the population weighted mean given by (\ref{equation popweighted}), while the aggregate model (\ref{equation aggregate}) uses the populations to weight the risks. Inference for all the Bayesian models (that is not \textbf{Model-GLM}) was based on 120,000 McMC samples, which were generated from 6 parallel Markov chains, and convergence was visually checked by examining trace plots of sample parameters including the pollution-health relationship. The main study results are presented in the next subsection, while the following subsection presents some additional sensitivity analyses.

\subsection{Main results}
The estimated relationships between each pollutant and respiratory hospital admissions from each model are displayed in the top panel of  Table \ref{results1}, and are presented as relative risks for a realistic increase in pollution concentrations. These increases are 5$\mu gm^{-3}$ for NO$_{2}$ and 1$\mu gm^{-3}$ for PM$_{2.5}$ and PM$_{10}$, because NO$_{2}$ has larger values and hence larger variation across  England. Overall, the table shows evidence of substantial relationships between air pollution and respiratory disease risk, as 17 of the 18 estimated relative risks have 95$\%$ uncertainty intervals that do not include the null risk of one. The most definitive effects are observed for NO$_{2}$ and PM$_{2.5}$, while weaker effects are observed for PM$_{10}$. \\

The table also shows clear evidence that the model used to allow for residual spatial autocorrelation can have a large impact on health effects estimation, with relative risks for NO$_{2}$ ranging between a 8.5$\%$ and a 9.9$\%$ increase in disease risk for a 5$\mu gm^{-3}$ increase in concentrations. The results for PM$_{2.5}$ and PM$_{10}$ are even more striking, with estimated risks ranging between a 3.3$\%$ and a 6.5$\%$ increase for PM$_{2.5}$ and between 0.9$\%$ and 3.3$\%$ for PM$_{10}$. The exploratory analysis presented in Section 2 suggests that the residuals after allowing for the known covariates are locally and not globally spatially smooth (see the bottom right panel of Figure \ref{data1}), and a comparison between \textbf{Model-CAR} and \textbf{Model-Local} suggests that wrongly fitting a globally smooth CAR model has resulted in substantial overestimation of the pollution-health effects in this study. Finally, a comparison of \textbf{Model-Local}  and \textbf{Model-Local-Agg} shows that ignoring within-area variation in the pollution concentrations has no effect on the results, which was suggested by the simulation study as the estimated effect sizes ($\hat{\alpha}$) are relatively small. NO$_{2}$ shows the strongest mean-variance relationships within a Local Authority (highest correlation coefficients in Section 2),  and even for this pollutant the bias term, assuming linearity and normality (see Section 3.3), $0.5b\alpha^{2}$  is just 0.000122.

\subsection{Sensitivity analyses}
We then undertook a sensitivity analysis to assess the impact of changing the model assumptions of \textbf{Model-Local-Agg}, and the results are displayed in the bottom panel of Table \ref{results1}.  This analysis included: (i) changing the maximum number of risk classes $G$ from 5 to 3; (ii) changing the hyperparameters $(a, b)$  of the inverse-gamma distribution for $\tau^{2}$ from $(a=0.001, b=0.001)$ to 
 $(a=0.1, b=0.1)$ and  $(a=0.5, b=0.0005)$; and (iii) removing the population weighting in (\ref{equation aggregate}) to give equal weight to all pollution concentrations. The table shows that changing $G$ and $(a, b)$ had almost no impact on the results, as the estimated relative risks for all pollutants remained almost identical.  Removing the population weighting also had little effect, although for NO$_{2}$ the estimated risk dropped by 0.6$\%$ from 1.087 to 1.081.

\begin{table}
\small\caption{Estimated relative risks and 95$\%$ uncertainty intervals (confidence intervals for \textbf{Model-GLM} and credible intervals for the remaining models) for a 5$\mu gm^{-3}$ (NO$_{2}$) and a 1$\mu gm^{-3}$ (PM$_{2.5}$ and PM$_{10}$) increase in pollution concentrations from the models considered in this paper.}\label{results1}
\small\centering\begin{tabular}{lrrr}
\hline
\raisebox{-1.5ex}[0pt]{\textbf{Model}}&\multicolumn{3}{c}{\textbf{Pollutant}}\\
&\textbf{NO$_{2}$} &\textbf{PM$_{2.5}$} &\textbf{PM$_{10}$}\\  \hline

\textbf{Main results}&&&\\
Model-GLM&1.085     (1.053,    1.119)& 1.033     (1.006,     1.061) &1.009     (0.990,     1.028) \\
Model-CAR&1.092 (1.050, 1.137)&1.065 (1.029, 1.103) &1.033 (1.005, 1.063)\\
Model-Local&1.086 (1.065, 1.106)&1.037 (1.026, 1.049)&1.012 (1.003, 1.022)\\
Model-Local-Agg&1.087 (1.072, 1.102)&1.037 (1.026, 1.048) &1.013 (1.005, 1.021)\\
Model-HH&1.089 (1.087, 1.091) &1.047 (1.045, 1.048) &1.019 (1.018, 1.021)\\
Model-LM&1.099      (1.084,      1.113)&1.047      (1.034,      1.061)&1.033     (1.023,      1.044) \\\hline

\textbf{Sensitivity analysis}&&&\\
G=3&1.090 (1.072, 1.107)& 1.037 (1.022, 1.055) &1.010 (0.999, 1.025)\\
a=0.1, b=0.1&1.088 (1.069, 1.110) &1.037 (1.024, 1.050)& 1.014 (1.004, 1.025)\\
a=0.5, b=0.0005&1.086 (1.071, 1.101)& 1.034 (1.020, 1.047)& 1.012 (1.002, 1.023)\\
No population weighting& 1.081 (1.065, 1.096)& 1.035 (1.023, 1.048)& 1.013 (1.004, 1.021)\\\hline
\end{tabular}
\normalsize
\end{table}

\section{Discussion}
This paper is the first to propose an integrated modelling framework for estimating the long-term effects of air pollution on human health that accounts for localised spatial  autocorrelation in the disease data and the inherent spatial misalignment between the exposure and the response. The model proposed here will be widely applicable to geographical association studies beyond the air pollution arena, and is available for other researchers to use via the R package \emph{CARBayes} located at \emph{http://www.R-project.org/}. This paper also provides an in-depth simulation study into the impact of spatial autocorrelation and spatial misalignment on fixed effects estimation, and presents a new study into the long-term effects of air pollution in England in 2010.\\

One of the main findings of this paper is that inappropriate control for residual (i.e. after the effects of known covariates have been removed) spatial autocorrelation in the disease data can lead to incorrect fixed effects estimation. This problem encompasses both point estimation and uncertainty quantification, and the first simulation study presented here shows some interesting results. Firstly, if, as is the case with the respiratory admissions data presented here, the residual spatial autocorrelation is not globally smooth, then wrongly assuming it is leads to substantially poorer estimation of covariate effects  (in terms of RMSE) compared with using a localised smoothing model. Differences in fixed effects estimates were also seen empirically in the real data in Section 5, although one is of course unable to say which estimate is `correct' in this case. The other main finding from the simulation study is that as expected the greater the level of confounding between the fixed effects and the unmeasured spatial structure, the poorer all models do in terms of fixed effects estimation. This poor performance again encompasses point estimation and uncertainty quantification. However, what is surprising is that the commonly used CAR model, which has been subject to recent criticisms by \cite{reich2006} and others, does no worse than other more sophisticated models, and in fact outperforms the orthogonal smoothing model proposed by \cite{hughes2013}.\\

The other main finding of this paper is that for air pollution and health studies, where effect sizes are typically small,  ignoring the within-area variation in the exposure caused by the spatial misalignment of the data does not lead to systematic bias. This result is illustrated in the second simulation study presented in Section 4, and is corroborated by the real data results presented in Section 5. The magnitude of any such bias depends on a number of factors, including the effect size being estimated, the distributional shape of the within-area variation in the exposure, and the relationship between the mean and higher order moments of the within-area exposure distribution. Here we only considered a Gaussian distribution for the exposures which is completely characterised by its first two moments and is amenable to analytic study, but if the distribution is skewed then alternatives such as a log-normal distribution may be appropriate instead.  However, in general within-area variation in an exposure should not be averaged away by computing a mean,  as bias can result if the above conditions are right as shown in the simulation study in Section 4.\\

The spatial air pollution and disease data used in this study are routinely available for multiple consecutive time periods, and in future work we wish to extend the model proposed here to the spatio-temporal domain. Furthermore, the modelled concentrations used here have complete spatial coverage but are modelled estimates rather than measured concentrations, and as such are prone to biases and uncertainties. However no information on these are available, and thus we plan to develop a two-stage disease model, where the first stage provides better estimates of pollution by fusing the modelled concentrations with observed monitor data using techniques  similar to \cite{berrocal2009}.

\section*{Acknowledgements}
This work was funded by the UK Engineering and Physical Sciences Research Council (EPSRC), via grant EP/J017442/1.

\bibliographystyle{chicago}
\bibliography{Lee}

\begin{thebibliography}{}

\bibitem[\protect\citeauthoryear{{AEA}}{{AEA}}{2011}]{AEA2011}
{AEA} (2011).
\newblock {UK} modelling under the {A}ir {Q}uality {D}irective (2008/50/ec) for
  2010 covering the following air quality pollutants: {SO}$_{2}$, {NO}$_{x}$,
  {NO}$_{2}$, {PM}$_{10}$, {PM}$_{2.5}$, lead, benezene, {CO} and ozone.
\newblock Technical report.

\bibitem[\protect\citeauthoryear{Bell and Davis}{Bell and
  Davis}{2001}]{bell2001}
Bell, M. and D.~Davis (2001).
\newblock Reassessment of the {L}ethal {L}ondon {F}og of 1952: {N}ovel
  {I}ndicators or {A}cute and {C}hronic {C}onsequences of {A}cute {E}xposure to
  {A}ir {P}ollution.
\newblock {\em Environmental {H}ealth {P}erspectives\/}~{\em 109}, 389--394.

\bibitem[\protect\citeauthoryear{Berrocal, Gelfand, and Holland}{Berrocal
  et~al.}{2009}]{berrocal2009}
Berrocal, V., A.~Gelfand, and D.~Holland (2009).
\newblock A {S}patio-{T}emporal {D}ownscaler for {O}utput {F}rom {N}umerical
  {M}odels.
\newblock {\em Journal of {A}gricultural, {B}iological and {E}nvironmental
  {S}tatistics\/}~{\em 15}, 176--197.

\bibitem[\protect\citeauthoryear{Besag, York, and Mollie}{Besag
  et~al.}{1991}]{besag1991}
Besag, J., J.~York, and A.~Mollie (1991).
\newblock Bayesian image restoration with two applications in spatial
  statistics.
\newblock {\em Annals of the {I}nstitute of {S}tatistics and
  {M}athematics\/}~{\em 43}, 1--59.

\bibitem[\protect\citeauthoryear{Cesaroni, Forastiere, Stafoggia, Andersen,
  Badaloni, Beelen, Caracciolo, {de Faire}, Erbel, Eriksen, Fratiglioni,
  Galassi, Hampel, Heier, Hennig, Hilding, Hoffmann, Houthuijs, Jšckel, Korek,
  Lanki, Leander, Magnusson, Migliore, Ostenson, Overvad, Pedersen, Pekkanen,
  Penell, Pershagen, Pyko, {Raaschou-Nielsen}, Ranzi, Ricceri, Sacerdote,
  Salomaa, Swart, Turunen, Vineis, Weinmayr, Wolf, Hoek, Brunekreef, and
  Peters}{Cesaroni et~al.}{2014}]{cesaroni2014}
Cesaroni, G., F.~Forastiere, M.~Stafoggia, Z.~Andersen, C.~Badaloni, R.~Beelen,
  B.~Caracciolo, U.~{de Faire}, R.~Erbel, K.~Eriksen, L.~Fratiglioni,
  C.~Galassi, R.~Hampel, M.~Heier, F.~Hennig, A.~Hilding, B.~Hoffmann,
  D.~Houthuijs, K.~Jšckel, M.~Korek, T.~Lanki, K.~Leander, P.~Magnusson,
  E.~Migliore, C.~Ostenson, K.~Overvad, N.~Pedersen, J.~Pekkanen, J.~Penell,
  G.~Pershagen, A.~Pyko, O.~{Raaschou-Nielsen}, A.~Ranzi, F.~Ricceri,
  C.~Sacerdote, V.~Salomaa, V.~Swart, A.~Turunen, P.~Vineis, G.~Weinmayr,
  K.~Wolf, K~and{de Hoogh}, G.~Hoek, B.~Brunekreef, and A.~Peters (2014).
\newblock Long term exposure to ambient air pollution and incidence of acute
  coronary events: prospective cohort study and meta-analysis in 11 {E}uropean
  cohorts from the {ESCAPE} {P}roject.
\newblock {\em British {M}edical {J}ournal\/}~{\em 348}, f7412.

\bibitem[\protect\citeauthoryear{Clayton, Bernardinelli, and Montomoli}{Clayton
  et~al.}{1993}]{clayton1993}
Clayton, D., L.~Bernardinelli, and C.~Montomoli (1993).
\newblock Spatial {C}orrelation in {E}cological {A}nalysis.
\newblock {\em Int {J} {E}pidemiol\/}~{\em 22}, 1193--1202,
  DOI:10.1093/ije/22.6.1193.

\bibitem[\protect\citeauthoryear{Dockery, Pope, Xu, Spengler, Ware, Fay,
  Ferris, and Speizer}{Dockery et~al.}{1993}]{dockery1993}
Dockery, D., C.~Pope, X.~Xu, J.~Spengler, J.~Ware, M.~Fay, B.~Ferris, and
  F.~Speizer (1993).
\newblock An {A}ssociation {B}etween {A}ir {P}ollution {A}nd {M}ortality {I}n
  {S}ix {U.S.} {C}ities.
\newblock {\em The {N}ew {E}ngland {J}ournal of {M}edicine\/}~{\em 329},
  1753--1759.

\bibitem[\protect\citeauthoryear{Elliott, Shaddick, Wakefield, Hoogh, and
  Briggs}{Elliott et~al.}{2007}]{elliott2007}
Elliott, P., G.~Shaddick, J.~Wakefield, C.~Hoogh, and D.~Briggs (2007).
\newblock Long-term associations of outdoor air pollution with mortality in
  {G}reat {B}ritain.
\newblock {\em Thorax\/}~{\em 62}, 1088--1094.

\bibitem[\protect\citeauthoryear{Greven, Dominici, and Zeger}{Greven
  et~al.}{2011}]{greven2011}
Greven, S., F.~Dominici, and S.~Zeger (2011).
\newblock An {A}pproach to the {E}stimation of {C}hronic {A}ir {P}ollution
  {E}ffects {U}sing {S}patio-{T}emporal {I}nformation.
\newblock {\em Journal of the {A}merican {S}tatistical {A}ssociation\/}~{\em
  106}, 396--406.

\bibitem[\protect\citeauthoryear{Haining, Li, Maheswaran, Blangiardo, Law,
  Best, and Richardson}{Haining et~al.}{2010}]{haining2010}
Haining, R., G.~Li, R.~Maheswaran, M.~Blangiardo, J.~Law, N.~Best, and
  S.~Richardson (2010).
\newblock Inference from ecological models: estimating the relative risk of
  stroke from air pollution exposure using small area data.
\newblock {\em Spatial {S}patio-temporal {E}pidemiology\/}~{\em 1}, 123--131.

\bibitem[\protect\citeauthoryear{Hughes and Haran}{Hughes and
  Haran}{2013}]{hughes2013}
Hughes, J. and M.~Haran (2013).
\newblock Dimension reduction and alleviation of confounding for spatial
  generalized linear mixed models.
\newblock {\em Journal of the {R}oyal {S}tatistical {S}ociety {S}eries
  {B}\/}~{\em 75}, 139--159.

\bibitem[\protect\citeauthoryear{Janes, Dominici, and Zeger}{Janes
  et~al.}{2007}]{janes2007}
Janes, H., F.~Dominici, and S.~Zeger (2007).
\newblock Trends in {A}ir {P}ollution and {M}ortality: {A}n {A}pproach to the
  {A}ssessement of {U}nmeasured {C}onfounding.
\newblock {\em Epidemiology\/}~{\em 18}, 416--423.

\bibitem[\protect\citeauthoryear{Jerrett, Buzzelli, Burnett, and
  DeLuca}{Jerrett et~al.}{2005}]{jerrett2005}
Jerrett, M., M.~Buzzelli, R.~Burnett, and P.~DeLuca (2005).
\newblock Particulate air pollution, social confounders, and mortality in small
  areas of an industrial city.
\newblock {\em Soc {S}ci {M}ed\/}~{\em 60}, 2845--2863.

\bibitem[\protect\citeauthoryear{Lawson, Choi, Cai, Hossain, Kirby, and
  Liu}{Lawson et~al.}{2012}]{lawson2012}
Lawson, A., J.~Choi, B.~Cai, M.~Hossain, R.~Kirby, and J.~Liu (2012).
\newblock Bayesian 2-{S}tage {S}pace-{T}ime {M}ixture {M}odeling {W}ith
  {S}patial {M}isalignment of the {E}xposure in {S}mall {A}rea {H}ealth {D}ata.
\newblock {\em J {A}gr, {B}iol {E}nvir {S}\/}~{\em 17}, 417--441.

\bibitem[\protect\citeauthoryear{Lee, Ferguson, and Mitchell}{Lee
  et~al.}{2009}]{lee2009}
Lee, D., C.~Ferguson, and R.~Mitchell (2009).
\newblock Air pollution and health in {S}cotland: a multicity study.
\newblock {\em Biostatistics\/}~{\em 10}, 409--423.

\bibitem[\protect\citeauthoryear{Lee and Mitchell}{Lee and
  Mitchell}{2013}]{lee2013}
Lee, D. and R.~Mitchell (2013).
\newblock Locally adaptive spatial smoothing using conditional autoregressive
  models.
\newblock {\em Journal of the {R}oyal {S}tatistical {S}ociety {S}tries
  {C}\/}~{\em 62}, 593--608.

\bibitem[\protect\citeauthoryear{Lee, Rushworth, and Sahu}{Lee
  et~al.}{2014}]{lee2014}
Lee, D., A.~Rushworth, and S.~Sahu (2014).
\newblock A {B}ayesian {L}ocalised {C}onditional {A}utoregressive {M}odel for
  {E}stimating the {H}ealth {E}ffecs of {A}ir {P}ollution.
\newblock {\em Biometrics\/}~{\em 70}, 419--429.

\bibitem[\protect\citeauthoryear{Leroux, Lei, and Breslow}{Leroux
  et~al.}{1999}]{leroux1999}
Leroux, B., X.~Lei, and N.~Breslow (1999).
\newblock {\em Estimation of disease rates in small areas: {A} new mixed model
  for spatial dependence. \textit{Statistical {M}odels in {E}pidemiology, the
  {E}nvironment and {C}linical {T}rials, Halloran, M and Berry, D (eds)}}, pp.\
   135--178.
\newblock Springer-Verlag, {N}ew {Y}ork.

\bibitem[\protect\citeauthoryear{Lu, Reilly, Banerjee, and Carlin}{Lu
  et~al.}{2007}]{lu2007}
Lu, H., C.~Reilly, S.~Banerjee, and B.~Carlin (2007).
\newblock Bayesian areal wombling via adjacency modelling.
\newblock {\em Environ {E}col {S}tat\/}~{\em 14}, 433--452.

\bibitem[\protect\citeauthoryear{Maheswaran, Haining, Brindley, Law, Pearson,
  Fryers, Wise, and Campbell}{Maheswaran et~al.}{2005}]{maheswaran2005}
Maheswaran, R., R.~Haining, P.~Brindley, J.~Law, T.~Pearson, P.~Fryers,
  S.~Wise, and M.~Campbell (2005).
\newblock Outdoor air pollution and stroke in {S}heffield, {U}nited {K}ingdom.
\newblock {\em Stroke\/}~{\em 36}, 239--243.

\bibitem[\protect\citeauthoryear{{R Core Team}}{{R Core Team}}{2013}]{R}
{R Core Team} (2013).
\newblock {\em R: A Language and Environment for Statistical Computing}.
\newblock Vienna, Austria: R Foundation for Statistical Computing.

\bibitem[\protect\citeauthoryear{Reich, Hodges, and Zadnik}{Reich
  et~al.}{2006}]{reich2006}
Reich, B., J.~Hodges, and V.~Zadnik (2006).
\newblock Effects of {R}esidual {S}moothing on the {P}osterior of the {F}ixed
  {E}ffects in {D}isease-{M}apping {M}odels.
\newblock {\em Biometrics\/}~{\em 62}, 1197--1206.

\bibitem[\protect\citeauthoryear{Wakefield and Salway}{Wakefield and
  Salway}{2001}]{wakefield2001}
Wakefield, J. and R.~Salway (2001).
\newblock A statistical framework for ecological and aggregate studies.
\newblock {\em Journal of the {R}oyal {S}tatistical {S}ociety {S}eries
  {A}\/}~{\em 164}, 119--137.

\bibitem[\protect\citeauthoryear{Wakefield and Shaddick}{Wakefield and
  Shaddick}{2006}]{wakefield2006}
Wakefield, J. and G.~Shaddick (2006).
\newblock Health exposure modelling and the ecological fallacy.
\newblock {\em Biostatistics\/}~{\em 7}, 438--455.

\bibitem[\protect\citeauthoryear{Young, Gotway, Yang, Kearney, and
  {DuClos}}{Young et~al.}{2009}]{young2009}
Young, L., C.~Gotway, J.~Yang, G.~Kearney, and C.~{DuClos} (2009).
\newblock Linking health and environmental data in geographical analysis:
  {I}t's so much more than centroids.
\newblock {\em Spatial {S}patio-temporal {E}pidemiology\/}~{\em 1}, 73--84.

\end{thebibliography}

\end{document}